\documentclass[twocolumn,showpacs,preprintnumbers,amsmath,amssymb]{revtex4}

\usepackage{array}
\usepackage{booktabs}
\usepackage{tabu}
\usepackage{dcolumn}
\usepackage{amsmath}
\usepackage{amsfonts}
\usepackage{amssymb}
\usepackage{graphicx}
\usepackage{subfigure}
\usepackage{graphicx}
\usepackage{dcolumn}
\usepackage{bm}
\usepackage{xcolor}

\def\be{\begin{equation}}
  \def\ee{\end{equation}}
\def\bea{\begin{eqnarray}}
\def\eea{\end{eqnarray}}
\def\f{\frac}
\def\n{\nonumber}
\def\l{\label}
\def\p{\phi}
\def\o{\over}
\def\R{\rho}
\def\pa{\partial}
\def\om{\omega}
\def\na{\nabla}
\def\P{\Phi}

\begin{document}

\title{Tsallis holographic dark energy for inflation}

\author{Abolhassan Mohammadi$^1$}
\email{a.mohammadi@uok.ac.ir; abolhassanm@gmail.com}
\author{Tayeb Golanbari$^1$}
\email{t.golanbari@uok.ac.ir; t.golanbari@gmail.com}
\author{Kazuharu Bamba$^2$}
\email{bamba@sss.fukushima-u.ac.jp}
\author{Iarley P. Lobo$^{3,4}$}
\email{iarley_lobo@fisica.ufpb.br}

\affiliation{
$^1$Department of Physics, Faculty of Science, University of Kurdistan,  Sanandaj, Iran.\\
$^2$Division of Human Support System, Faculty of Symbiotic Systems Science, Fukushima University, Fukushima 960-1296, Japan. \\
$^3$Department of Chemistry and Physics, Federal University of Para\'iba, Rodovia BR 079 - Km 12, 58397-000 Areia-PB,  Brazil. \\
$^4$Physics Department, Federal University of Lavras, Caixa Postal 3037, 37200-900 Lavras-MG, Brazil.
}
\date{\today}

\def\be{\begin{equation}}
  \def\ee{\end{equation}}
\def\bea{\begin{eqnarray}}
\def\eea{\end{eqnarray}}
\def\f{\frac}
\def\n{\nonumber}
\def\l{\label}
\def\p{\phi}
\def\o{\over}
\def\R{\rho}
\def\pa{\partial}
\def\om{\omega}
\def\na{\nabla}
\def\P{\Phi}

\begin{abstract}
The application of holographic principle in very early time is studied. The consideration of the principle in the late-time evolution will be a good motivation to study its role at the time of inflation. Since the length scale is expected to be small during inflation, the resulted energy density form the holographic principle is expected to be large enough to drive inflation. The entropy of the system is the main part of the holographic principle, in which modifying entropy will lead to a modified energy density. Here, instead of the original entropy, we are going to apply a modified entropy, known as Tsallis entropy which includes quantum corrections. The length scale is assumed to be GO length scale. Finding an analytical solution for the model, the slow-roll parameters, scalar spectral index, and tensor-to-scalar ratio are calculated. Comparing the result, with Planck $r-n_s$ diagram, we could find a parametric space for the constants of the model. Then, a correspondence between the holographic dark energy and the scalar field is constructed, and the outcome potentials are investigated. In the final part of the work, we have considered the recently proposed trans-Planckian censorship conjecture for the model.
\end{abstract}
\maketitle

\section{Introduction}
The inflationary scenario has received tremendous observational support during past couple of decades and it has become the cornerstone of any cosmological model. The scenario describes a very short extreme expansion for the universe at the very early times. Since the first introduction of inflation
\cite{starobinsky1980new,Guth:1980zm,albrecht1982cosmology,linde1982new,linde1983chaotic}, the scenario has received huge interest and it has been studied and modified in many different aspects \cite{Barenboim:2007ii,Franche:2010yj,Unnikrishnan:2012zu,Rezazadeh:2014fwa,Saaidi:2015kaa,
Fairbairn:2002yp,Mukohyama:2002cn,Feinstein:2002aj,Padmanabhan:2002cp,Aghamohammadi:2014aca,
Spalinski:2007dv,Bessada:2009pe,Weller:2011ey,Nazavari:2016yaa,
maeda2013stability,abolhasani2014primordial,alexander2015dynamics,tirandari2017anisotropic,
maartens2000chaotic,golanbari2014brane,Mohammadi:2020ake,Mohammadi:2020ctd,
berera1995warm,berera2000warm,hall2004scalar,Sayar:2017pam,Akhtari:2017mxc,Sheikhahmadi:2019gzs,Rasheed:2020syk,
Mohammadi:2018oku,Mohammadi:2019dpu,Mohammadi:2018zkf,Mohammadi:2019qeu,Mohammadi:2020ftb}. Inflation is usually assumed to be driven by a scalar field which is based on the slow-rolling assumptions \cite{Linde:2000kn,Linde:2005ht,Linde:2005vy,Linde:2004kg,Riotto:2002yw,Baumann:2009ds,Weinberg:2008zzc,Lyth:2009zz}. \\
The present accelerated expansion of the universe is associated to dark energy. Although the true nature of dark energy is not realized yet, there are many different candidates (refer to \cite{Bamba:2012cp} for a review on different models of dark energy). The scalar fields model are one of these candidates. The holographic dark energy (HDE) is another candidate of dark energy \cite{Hsu:2004ri,Horvat:2004vn,Li:2004rb}. The HDE is established based on the holographic principle \cite{tHooft:1993dmi} which is originated from the thermodynamics of black hole. The principle was even extended to the string theory by Susskind \cite{Susskind:1994vu}. The holographic principle states that the entropy of a system is measured with its area, rather than its volume \cite{tHooft:1993dmi,Susskind:1994vu,Witten:1998qj,Bousso:2002ju}. The formation of black hole puts out a limit which provides a connection between ultra-violet cutoff (short distance cutoff $\Lambda$) and the infrared cutoff (long distance cutoff $L$) \cite{Cohen:1998zx}. \\
Huge amount of researches are devoted to the application of HDE in late-time evolution of the universe and its role as dark energy \cite{Nojiri:2019skr,Nojiri:2019itp,Nojiri:2005pu,Nojiri:2020wmh} (for a review on HDE refer to \cite{Li:2004rb}). Then, it would be a good motivation to construct inflation based on the same energy density, i.e. HDE. The HDE is related to the inverse squared of the infrared cutoff. Since, the length scale is small in the inflationary times, it is expected to have a large energy density, enough to drive inflation. The infrared cutoff is related to the casuality. Due to this, it is taken as a form of horizon such as particle horizon, future event horizon, or Hubble length. Granda-Oliveros (GO) is known as another type of cutoff which was introduced in \cite{Granda:2008dk,Granda:2008tm} mostly based on dimensional reasons. This cutoff is a combination of the Hubble parameter and its time derivative. \\
The original HDE is based on the standard entropy, $S = A/4$. Including the quantum corrections, the area law could be modified in different types in which the Logarithmic corrections \cite{Mann:1996ze,Rovelli:1996dv,Ashtekar:1997yu,Kaul:2000kf,Das:2001ic}, arising from loop quantum gravity, and power-law correction \cite{Das:2008sy,Das:2010su,Das:2007mj,Radicella:2010ss}, arising in dealing with the entanglement of quantum fields, could be named as some examples of these quantum corrections. Another modification comes out from the fact that the gravitational and cosmological systems, which have divergency in the partition function, could not be described by Boltzmann-Gibbs theory. Rather, the thermodynamical entropy of such a system must be modified to the non-additive entropy, instead of using the additive one \cite{Tsallis:1987eu,Lyra:1998wz,Tsallis:1998ws,Wilk:1999dr}. Based on this argument, Tsallis and Cirto have shown that the entropy of black hole should be generalized to $S = \gamma A^\delta$ \cite{Tsallis_2013}, where $A$ is the area of the black hole horizon, $\delta$ is called the Tsallis parameter or nonextensive parameter, and $\gamma$ is an unknown parameter. For $\gamma = 1/4$ and $\delta = 1$ it returns to the BG entropy. The power-law function of the entropy, which has been inspired by Tsallis entropy, is suggested by quantum gravity investigations \cite{Rashki:2014noa,Tavayef:2018xwx}. \\
The Tsallis entropy inspires a modified energy density. The effect of the energy density for the late-time evolution of the universe has been studied for different types of the infrared cutoffs
\cite{Tavayef:2018xwx,Jahromi:2018xxh,Sharif:2019seo,Saridakis:2018unr,Ghaffari:2019qcv,Ghaffari:2018wks,
Sheykhi:2018dpn,Mamon:2020wnh}. \\
During the present work, we are going to consider the resulted energy density of Tsallis entropy as the source of inflation in which the infrared cutoff is picked out to be the GO cutoff. Inflation for the standard HDE has been considered for different cutoffs in \cite{Nojiri:2019kkp,Oliveros:2019rnq,Chakraborty_2020}. Following the assumption that the quantum correction already has been applied to the entropy, the correction in the infrared cutoff is not taken account. An analytical solution for the Hubble parameter is obtained which is utilized to compute the slow-roll parameters, scalar spectral index, and also the tensor-to-scalar ratio. Applying some mathematica coding, the results of the model are compared with the Planck $r-n_s$ diagram and parametric space for the free constants of the model is derived, in which for every point in the space the model comes to a good agreement with data. Using the resulted values for the constant, we then establish a correspondence between the Tsallis HDE and the scalar field to construct a potential. In the final part, we investigate the validity of the trans-Planckian censorship conjecture (TCC) \cite{Bedroya:2019snp,Bedroya:2019tba,Brandenberger:2019eni,Lin:2019pmj} for the model. Although the recently proposed TCC is a theoretical conjecture, there is a rising expectation for the inflationary model to satisfy the conjecture, besides the observational data.

\section{Tsallis holographic dark energy}
The derivation of original holographic dark energy (HDE), which is presented as $\rho = 3 c^2 M_p^2 / L^2$ (where $c$ is the speed of light in vacuum and $M_p$ is the Planck mass), is based on the well-known entropy-area relation $S = A/4$ of black hole where $A$ is the area of the horizon. It is understandable that any modification to the entropy-area relation will lead to a modified HDE. Tsallis and Cirto have shown that by considering the quantum correction the entropy-area elation is modified as \cite{Tsallis_2013}
\begin{equation}\label{TsallisS}
  S = \gamma A^\delta,
\end{equation}
where $\gamma$ in an unknown constant and $\delta$ is the non-additivity parameter \cite{Tsallis_2013,Tavayef:2018xwx,Jahromi:2018xxh,Sharif:2019seo,Saridakis:2018unr,Ghaffari:2019qcv,Ghaffari:2018wks,
Sheykhi:2018dpn,Mamon:2020wnh}. For $\gamma = 1/ G$ and $\delta = 1$, the Bekenstein entropy is recovered.
Scaling the number of degrees of freedom of a physical system with its bounding area is known as the holography principle \cite{tHooft:1993dmi}. By considering an Infrared cutoff, Cohen \textit{et al.} have established a relation between the entropy (S), IR cutoff (L), and the UV cutoff ($\Lambda$) as follows
\begin{equation*}
  L^3 \Lambda^3 \leq S^{3/4}.
\end{equation*}
Substituting the entropy from Eq.\eqref{TsallisS}, and taking the area as $A = 4\pi L^2$, one arrives at
\begin{equation}\label{TsallisLambda}
  \Lambda^4 \leq \gamma \; (4\pi)^\delta \; L^{2\delta - 4}\, ,
\end{equation}
in which $\Lambda^4$ indicates the vacuum energy density. Based on the HDE hypothesis, $\Lambda^4$ is taken as the energy density of dark energy ($\rho_{de}$). Then, following Eq.\eqref{TsallisLambda}, the modified energy density is obtained which is named as the Tsallis HDE (TDH), given by
\begin{equation}\label{THDErho}
  \rho_{THDE} = B c^2 \; L^{2\delta - 4},
\end{equation}
where the unknown parameter $B$ is defined as $B \equiv \gamma \; (4\pi)^\delta c^2$ with dimension ${\rm M^{2\delta}}$. \\
Assuming a spatially flat FLRW metric as the geometry of the spacetime, the Friedmann equations are given by
\begin{eqnarray}\label{Friedmann}
  H^2 & = & {B c^2 \over 3M_p^2} \; L^{2\delta - 4} + {1 \over 3M_p^2} \; \rho_m, \nonumber \\
  2\dot{H} + 3H^2 & = & {-1 \over M_p^2} \; p_{THDE}.
\end{eqnarray}
where $\rho_m$ is the matter density, and $p_{THDE}$ is the HDE pressure. Ignoring the interaction between dark energy and other possible components, we have a conservation equation for TDHE as
\begin{equation}\label{TDHEconservation}
  \rho_{TDHE} + \; 3H (1 + \omega_{THDE}) \; \rho_{THDE} = 0\, ,
\end{equation}
in which the equation of state parameter $\omega_{THDE}$ is read as $\omega_{THDE} = p_{THDE} / \rho_{THDE}$, which could be read from Eq.\eqref{Friedmann} as
\begin{equation}\label{omega}
  \omega_{THDE} = -1 - 2M_p^2 \; {\dot{H} \over \rho_{THDE}} = -1 - {2M_p^2 \over B c^2} \; {\dot{H} \over L^{2\delta - 4}} .
\end{equation}
There are different choices for the IR cutoff $L$. The simplest choice is the Hubble length, $H^{-1}$, and the other choices are particle horizon and future event horizon. Another candidate of the IR cutoff is the GO cutoff, proposed in \cite{Granda:2008dk,Granda:2008tm} with dimensional motivation. The length scale in general is a combination of the Hubble parameter and its time derivative
\begin{equation}\label{RicciIR}
  L^{-2}_{IR} = \alpha H^2 + \beta \dot{H}.
\end{equation}
where $\alpha$ and $\beta$ are two dimensionless constant. As mentioned in the introduction, the Tsallis entropy has been raised from some quantum corrections and already encodes quantum gravitational effects. Then, it is assumed that the correction have already made in the entropy, and the GO cutoff are not required to be modified due to the high energy regime.

\section{inflation driven by THDE}
Inflation is known as a period of accelerated expansion phase of the universe at very early time. This acceleration phase is supported by a dark energy which for inflation it is usually taken as a scalar field model. Here, it is assumed that the expansion phase of the universe is provided by THDE with energy density \eqref{THDErho} and the GO length scale \eqref{RicciIR} as IR cutoff. Due to the rapid expansion, the contribution of other component is ignored, and the Friedmann equation is given by
\begin{equation}\label{FriedmannInf}
  H^2 = {B c^2 \over 3 M_p^2} \; \Big( \alpha H^2 + \beta \dot{H} \Big)^{2-\delta}\, .
\end{equation}
After some manipulation, one can extract the time derivative of the Hubble parameter
\begin{equation}\label{Hdot}
  \dot{H} = {H^2 \over \beta} \; \left[ \left( 3M_p^2 \over B c^2 \right)^{1 \over (2-\delta)} \; \Big( H^2 \Big)^{\delta -1 \over 2-\delta} - \alpha \right]\, .
\end{equation}
Change of variable $N=\ln\big( a / a_i \big)$ from which $dN = H dt$ simplifies the oncoming calculation (where $a_i$ is an initial value of the scale factor $a$). By this change, one has $\dot{H} = {1 \over 2} \; {dH^2 \over dN}$. Taking integration results in a Hubble parameter in terms of the number of e-folds
\begin{equation}\label{HN}
  \ln \left[ \tilde{H}^2 \left( C \; \Big( \tilde{H}^2 \Big)^{\delta - 1 \over 2-\delta} \right)^{\delta-2 \over \delta - 1} - \alpha \right] \; \Bigg|_{\tilde{H}_i}^{\tilde{H}_e} = {-2 \alpha N \over \beta}\, ,
\end{equation}
in which the constant $C$ s defined as
\begin{equation*}
  C \equiv \left( 3M_p^2 \over B c^2 \right)^{1 \over (2-\delta)} \; M_p^{2(\delta-1) \over 2-\delta},
\end{equation*}
and $\tilde{H}$ is named the dimensionless Hubble parameter given by $\tilde{H} \equiv H / M_p$. Of course, integration from Eq.\eqref{Hdot} with respect to the time gives $H$ versus time $t$.\\

\begin{figure}[h]
  \centering
  \subfigure[]{\includegraphics[width=6cm]{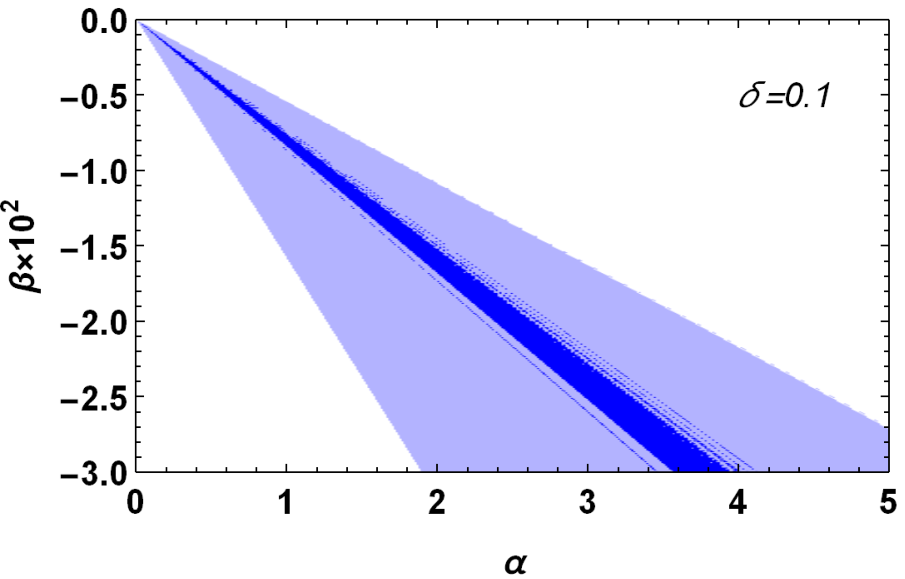}}
  \subfigure[]{\includegraphics[width=6cm]{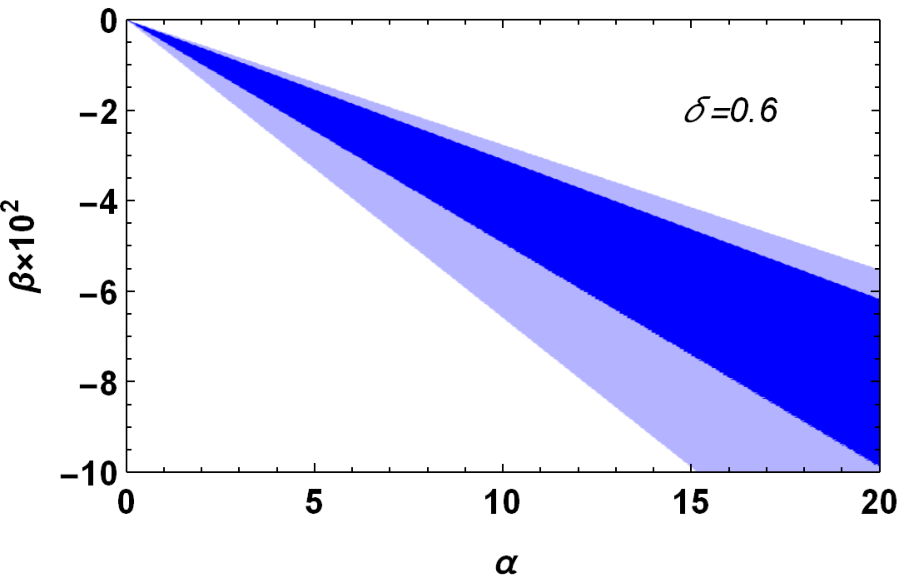}}
  \caption{The parametric space of ($\alpha$ , $\beta$) in which for every point in the range, the model comes to an agreement with data. To constrain the constants $\alpha$ and $\beta$, we have used the $r-n_s$ diagram of Planck-2018.}\label{alphabeta}
\end{figure}

\begin{figure}[h]
  \centering
  \includegraphics[width=6cm]{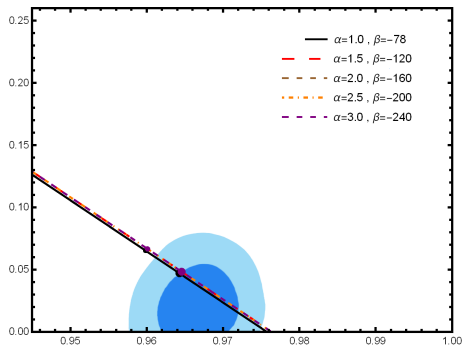}
  \caption{The $r-n_s$ curve of the model for five different ($\alpha$ , $\beta$) points, taken from Fig.\ref{alphabeta}, have been plotted. The $\alpha$ and $\beta$ points are picked put from the parametric space of Fig.\ref{alphabeta}.}\label{rnsdiagram}
\end{figure}

The slow-roll parameters, which are the essential parameters of the slow-roll inflation, are derived through the equation \eqref{Hdot}. Doing straightforward calculation, one obtains
\begin{equation}\label{epsilon1}
  \epsilon_1 = {- \dot{H} \over H} = {-1 \over \beta} \; \left[ C \; \Big( \tilde{H}^2 \Big)^{\delta - 1 \over 2-\delta} - \alpha \right]\, .
\end{equation}
The rest of the slow-roll parameters are defined through the usual definition as $\epsilon_{n+1} = d\ln(\epsilon_n)/dN$. Then, for the second one, we have
\begin{equation}\label{epsilon2}
  \epsilon_2 = {\dot{\epsilon}_1 \over H \epsilon_1} = {2 C \over \beta} \; \left( {\delta - 1 \over 2 - \delta} \right) \; \Big( \tilde{H}^2 \Big)^{\delta - 1 \over 2-\delta}.
\end{equation}
These parameters are assumed to be very small at the beginning of inflation. The main purpose is to obtained the main perturbation parameters at this time, and compare the result with observation and consider their consistency. Inflation ends when $\epsilon_1 = 1$. The Hubble parameter at this time is obtained as
\begin{equation}\label{Hend}
  \tilde{H_e^2} = \left( {C \over \alpha-\beta} \right)^{\delta - 2 \over \delta -1}.
\end{equation}
Then, from Eq.\eqref{HN}, the Hubble parameter is obtained at earlier time of inflation, including the horizon crossing time,
\begin{equation}\label{Hb2star}
  \tilde{H}_\star^2 = \left[ {C \over \alpha} \; \left( 1+ {\beta \over \alpha - \beta}\; e^{2\alpha N / \beta} \right) \right]^{\delta-2 \over \delta-1}.
\end{equation}
Substituting it in the slow-roll parameter, they will be calculated for earlier time as well. \\
Following \cite{Nojiri:2019kkp}, the usual perturbation procedure for deriving scalar spectral index, $n_s-1$, and tensor-to-scalar ratio, $r$, is picked out as the approximate approach instead of dealing with the perturbation analysis of HDE. The usual perturbation procedure is a good approximation here, which leads to
\begin{eqnarray}\label{nsr}
  n_s & = & 1 - 2\epsilon_1 - 2 \epsilon_2, \nonumber \\
  r & = & 16 \epsilon_1.
\end{eqnarray}
The above explained procedure is used to obtain the $n_s$ and $r$ at the time of horizon crossing, which indicates that they depend on the free constants of the model. Using the Planck $r-n_s$ diagram and applying some Mathematica code, one could find the range of the model constants in which for every point in the range, the model perfectly agrees with observational data. The, parametric space is presented in Fig.\ref{alphabeta}. \\
To test the result, five different $(\alpha,\beta)$ points are taken from the above parametric space, and $r-n_s$ curve of the model for the selected points have been plotted in Fig.\ref{rnsdiagram}, which indicates that the curves cross the $68\%$ regime.

The constant $C$ has no role in the slow-roll parameters at the time of horizon crossing. Therefore the scalar spectral index and the tensor-to-scalar ratio do not depend on $C$ at horizon crossing time. The constant is determined through the amplitude of the scalar perturbation, $\mathcal{P}_s = H^2 / 8\pi^2 \epsilon M_p^2$. The Planck data indicates that the amplitude of the observational data is of the order of $\mathcal{P}_s \propto 10^{-9}$. The constant $C$ then is obtained as
\begin{equation}\label{constantC}
  C = { \alpha \over \left( 1+ {\beta \over \alpha - \beta}\; e^{2\alpha N / \beta} \right) } \;
  \left( 8\pi^2 \epsilon \mathcal{P}_s \right)^{\delta-1 \over \delta-2}.
\end{equation}
Using the constants $\alpha$ and $\beta$ from the parametric space Fig.\ref{alphabeta}, the constant $C$ is calculated. Table \ref{Ctable} presents the value of the constant for different values of $\alpha$ and $\beta$.
\begin{table}
  \centering
  \begin{tabular}{p{0.8cm}p{1.2cm}p{2cm}}
    \hline
    $\ \alpha$ & $\quad \beta$ & $\qquad C$ \\[0.5mm]
    \hline
    $1.0$ & $\ -78$  & $4.76 \times 10^{-5}$ \\[1mm]
    $1.5$ & $-120$ & $7.30 \times 10^{-5}$ \\[1mm]
    $2.0$ & $-160$ & $9.73 \times 10^{-5}$ \\[1mm]
    $2.5$ & $-200$ & $1.21 \times 10^{-4}$ \\[1mm]
    $3.0$ & $-240$ & $1.46 \times 10^{-4}$ \\
    \hline
  \end{tabular}
  \caption{Estimating the values of the constant $C$ from the amplitude of the scalar perturbation, using the result of the parametric space of Fig.\ref{alphabeta}.}\label{Ctable}
\end{table}

\section{Correspondence between THDE and scalar field}
In this section, we show that it is possible to describe the behavior of inflation provided by the THDE approach into the dynamics of a scalar field in two different models.

\subsection{Canonical scalar field}
\begin{figure}[h]
  \centering
  \includegraphics[width=7cm]{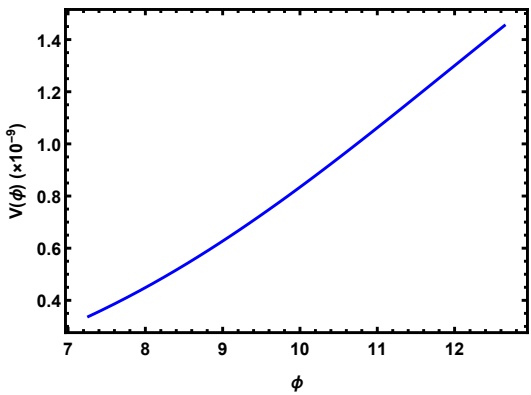}
  \caption{The constructed potential from the THDE for the canonical scalar field. The potential is plotted versus the scalar field. The constants are taken as $\alpha=1$, $\beta=-78$, and $\delta=0.1$.}\label{potcanonical}
\end{figure}

First we consider the correspondence between THDE and canonical self-interacting scalar field. The energy density and pressure of the scalar field is given by \cite{Bamba:2012cp,Yoo:2012ug,Novosyadlyj:2015zpa,Li:2011sd}
\begin{eqnarray}
  \rho_\phi &=& {1 \over 2} \; \dot{\phi}^2 + V(\phi), \nonumber \\
  p_\phi &=& {1 \over 2} \; \dot{\phi}^2 - V(\phi).
\end{eqnarray}
Establishing a correspondence between the THDE and the canonical scalar field, the potential is read as
\begin{equation}\label{pot}
  V(\phi) = \rho_{THDE} - {1 \over 2} \; \dot{\phi}^2 = \left( {1 - \omega_{THDE} \over 2} \right) \; \rho_{THDE} ,
\end{equation}
where we have used
\begin{equation}\label{phidot}
  \dot{\phi}^2 = \rho_{THDE} + p_{THDE} = \big( 1 + \omega_{THDE} \big) \; \rho_{THDE} .
\end{equation}
Substituting the $\omega_{THDE}$ from Eq.\eqref{omega} in \eqref{pot}, one arrives at
\begin{eqnarray}\label{potential}
  V(\phi) & = & \rho_{THDE} + M_p^2 \; \dot{H} \nonumber \\
         & = &  B c^2 \; \Big( \alpha H^2 + \beta \; \dot{H} \Big)^{2-\delta} + M_p^2 \; \dot{H} .
\end{eqnarray}
Inserting $\dot{H}$ from Eq.\eqref{Hdot}, the potential is obtained in terms of the Hubble parameter. The $\dot{\phi}$ follows the known equation $\dot{\phi}^2 = -2M_p^2 \dot{H}$ (which follows from the Friedmann equations). The time derivative of the scalar field could be rewritten as $\dot{\phi} = H \phi'$ in which the prime denotes derivative with respect to the number of e-folds, i.e. $\phi' =d\phi / dN$. Then, there is $\phi'^2 = -2M_p^2 \dot{H} / H^2$, and by using the definition of the first slow-roll parameter, the scalar field is obtained by taking the following integration
\begin{equation}\label{phi}
  \Delta \phi(N) = \sqrt{2} \; M_p \; \int_{0}^{N} \; \sqrt{{-1 \over \beta} \; \left[ C \; \Big( \tilde{H}^2 \Big)^{\delta - 1 \over 2-\delta} - \alpha \right]} \; dN ,
\end{equation}
where $N=0$ corresponds to the horizon crossing of perturbations. When solving the integral analytically one is faced with some difficulties mostly due to the presence of the incomplete gamma function. However, it could be solve numerically, and using the result in Eq.\eqref{potential}, one could illustrate the potential versus the scalar field as presented in Fig.\ref{potcanonical}.
It is realized that the scalar field rolls down from the top toward the minimum of the potential. To plot the potential, the Planck mass $M_p$ is taken as unity. The potential indicates that inflation occurs at the energy scale about $10^{-9}M_p^4 \; {\rm GeV^4}$, and the potential could be categorized as a large-field potential where the scalar field is bigger than the Planck mass.

\subsection{Tachyon field}
\begin{figure}[h]
  \centering
  \includegraphics[width=7cm]{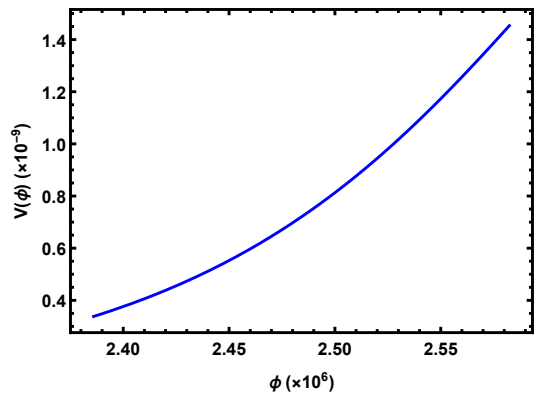}
  \caption{The constructed potential from THDE for the tachyon field is plotted versus the scalar field during inflation. The constants are taken as $\alpha=1$, $\beta=-78$, and $\delta=0.1$. }\label{potentialTachyon}
\end{figure}

The energy density and the pressure of the tachyon field is given by \cite{Padmanabhan:2002sh}
\begin{eqnarray}
  \rho_\phi &=& {V(\phi) \over \sqrt{1 - \dot{\phi}^2} } , \\
  p_\phi &=& - V(\phi) \; \sqrt{1 - \dot{\phi}^2} ,
\end{eqnarray}
and the equation of state of the field is read as $\omega_\phi = p_\phi / \rho_\phi = 1 - \dot{\phi}^2$. \\
By associating the energy density and pressure to the energy density and pressure of the THDE, the potential is obtained as
\begin{equation}\label{potTachyon}
  V(\phi) = \rho_{THDE} \; \sqrt{1 - \dot{\phi}^2}.
\end{equation}
Using the relation $\omega_\phi = \omega_{THDE}$, the time derivative of the scalar field is found as $\dot{\phi}^2 = 1 + \omega_{THDE}$. Using Eq.\eqref{omega}and the definition of the first slow-roll parameter, one arrives at
\begin{equation}\label{dphi}
 \phi'^2 = {2 \over 3 M_p^2} \; {\epsilon_1 \over \tilde{H}^2} ,
\end{equation}
which by taking integration, the change of the scalar field during inflation is obtained as
\begin{equation}\label{deltaphi}
  \Delta\phi = \sqrt{2 \over 3 M_p^2} \; \int_{0}^{N} \; \sqrt{{-1 \over \beta \tilde{H}^2 } \; \left[ C \; \Big( \tilde{H}^2 \Big)^{\delta - 1 \over 2-\delta} - \alpha \right]} \; dN .
\end{equation}
By numerically solving the integral and applying the result in Eq.\eqref{potTachyon}, the potential is obtained as a function of the scalar field. Fig.\ref{potentialTachyon} depicts the potential during inflation where it is shown that the scalar field rolls down from top toward the minimum.

\section{Trans-Planckian Censorship Conjecture}
The origin of the universe structure is the fluctuations in matter and energy. The casual mechanism for generating these fluctuations is provided by the inflationary scenario. The fluctuations produced during inflation are stretched out, cross the (Hubble) horizon, freeze and then come back to the horizon after inflation. The key point is that the produced fluctuations have a quantum origin, and as they cross the horizon, they behave classically. In late time, they enter the horizon and are probed in current cosmological observations. The crucial point is that if inflation lasted more than enough, it is possible to observe the length scale which would be originated on the scale smaller than the Planck length. This is known as the ``Trans-Planckian problem", see \cite{Bedroya:2019tba} for more information. How should we here treat the trans-Planckian mode? This is the question that does not arise in a consistent theory of quantum gravity. It is addressed as the ``Trans-Planckian Censorship Conjecture" (TCC).    \\
The TCC states that a length scale that crosses the horizon must not ever have had a length smaller than the Planck length \cite{Mohammadi:2020ctd,Bedroya:2019snp,Bedroya:2019tba,Brandenberger:2019eni,Lin:2019pmj}. The length scales never cross the horizon in the standard big bang cosmology, but in the inflationary scenario the story is different. The TCC is formulated as follow
\begin{equation}\label{TCC}
{l_p \over a_i } \;  <  \; {H_f^{-1} \over a_f},
\end{equation}
where $a_i$ and $a_f$ are respectively the universe scale factor at the beginning and end of inflation, $H_f$ is the Hubble parameter at the final time of inflation, and $l_p=m_p^{-1}$ is the Planck length. \\

Comparing the model with observational data, we could determine the constants of the model. Also, the Hubble parameter at the end and during inflation was specified. Using this result, one easily could examine the validity of the TCC, Eq.\eqref{TCC}. Using Eq.\eqref{Hend}, and after some manipulation, the condition could be rewritten as
\begin{equation}\label{TCCmode}
\left( {C \over \alpha - \beta} \right)^{\delta-2 \over \delta - 1} < \left( 8\pi \; e^N \right)^2
\end{equation}
The condition has been examined for different values of $\alpha$ and $\beta$ from the parametric space, Fig.\ref{alphabeta}. The result indicates that the LHS is of the order of $10^{-14}$ and the RHS is of the order of $10^{-54}$. It implies that the model does not satisfy the TCC. \\

Is there a trans-Planckian problem in inflation? The question has been discussed in detail in \cite{Dvali:2020cgt}. The TCC is all about tracking a perturbation back in time. The fluctuations are born in the inflationary time and cross the Hubble patch when they have a wavelength of the order of $L \sim H^{-1}$, due to the expansion of the universe. Then, by scaling back in time, one finds a wavelength shorter than the Planck length for enough inflation. However, is this a physically meaningful reasoning? The question has been answer in \cite{Dvali:2020cgt}. They argue that, in brief, to a certain initial time the fluctuations do not even exist. The great part of the de Sitter fluctuations are produced with wavelength $L \sim H^{-1}$, and only the fluctuations with wavelength $L \ll H^{-1}$ are exponentially suppressed, with factor $e^{-1/LH}$. Therefore, taking a wavelength back in time is a misleading point. Also, as the mode shrinks to scale beyond the Hubble patch,  the probability of materializing the de Sitter perturbation with trans-Planckian energy scale should be taken into account. The probability is suppressed by the factor $e^{-1/LH}$ as the mode shrinks to the Planck length. Taking into account this suppression indicates that the inflationary scenario might be free of the trans-Planckian problem. \\
Another point regarding the trans-Planckian regime, which is addressed in \cite{Dvali:2020cgt}, is the fuzziness that the definition carries on. It is stated that with the ordinary renormalizable theories with Wilsonian UV-completion one could probe an arbitrary short distance. However, in Einstein theory of gravity the tracking only goes on until one reaches the Planck length scale \cite{Dvali:2010bf}. It is shown that the minimal localization radius is described by a classical gravitational radius which turns out to be larger than the Compton wavelength, thus indicating that the described object is classical. It is an intrinsic feature of the Einstein gravity that by further tracking beyond the Planck scale, the theory classicalizes and presents a black hole \cite{Dvali:2010bf}. Therefore, by tracking perturbations to the trans-Planckian time, we are actually scaling them back to their classicalization. \\
Based on the above two points, the authors in \cite{Dvali:2020cgt} states that there is no trans-Planckian problem in inflationary scenario.

\section{Conclusion}
Even though an extensive amount of researches on the role of the holographic principle in explaining the late-time evolution of the universe, its application for the very early universe has recently been raised. In the presented work, the holographic principle was investigated for describing the inflationary scenario. The principle states that the energy density depends on the inverse of the length squared. Since the length scale is usually taken as the horizon, and it is believed that the horizon is decreasing during inflation, it is expected to have a large amount of energy to support inflation.\\
The HDE is originated from the entropy, which in the standard form linearly depends on the area, based on the holographic principle. However, the entropy could be modified by taking into account the quantum corrections. One of the modified entropies is known as Tsallis entropy which the corresponding energy density is assumed to be responsible for inflation. The length scale of the energy density is taken as the GO cutoff which is a combination of the Hubble parameter and its time derivative. \\
The equation was solved analytically and we found an exact solution for the Hubble parameter versus both time and number of e-folds. Applying the solution, the Hubble slow-roll parameters were derived. Then, the scalar spectral index and the tensor-to-scalar ratio are derived in terms of the model's constants. Comparing the model prediction about $n_s$ and $r$ with the $r-n_s$ diagram of Planck-2018, and using Mathematica coding, we found a parametric space for $\alpha$ and $\beta$ so that for every point in the space, the model has a perfect consistency with observational data. The results imply the ability of the Tsallis inflation for explaining the early universe.   \\
Next, we constructed a correspondence between THDE and two scalar field models as canonical and tachyon scalar fields. Using the $\alpha$ and $\beta$, obtained in the first part of the investigation, we could find the corresponding potential for each case. \\
The TCC was considered in the last part of the manuscript. It seems that there is a rising expectation for any inflationary model to satisfy this conjecture which imposes strong constraints on inflationary models. Using the obtained results for the free constant of the model, which were extracted by comparing with data, indicates that the model is far away from satisfying the conjecture. Then, although the model could be consistent with data, it is unable to satisfy the TCC.

\acknowledgements
The work of A.M. has been supported financially by ``Vice Chancellorship of Research and Technology, University of Kurdistan" under research Project No.99/11/19063. The work of T. G. has been supported financially by ``Vice Chancellorship of Research and Technology, University of Kurdistan" under research Project No.99/11/19305. The work of KB was partially supported by the JSPS KAKENHI Grant Number JP 25800136 and Competitive Research Funds for Fukushima University Faculty (19RI017). The work of IPL was partially supported by the National Council for Scientific and Technological Development - CNPq grant 306414/2020-1 and IPL would like to acknowledge the contribution of the COST Action CA18108.










\bibliography{THDEinflation}



\end{document}